\documentclass[oldversion,letter]{aa}  
\usepackage{graphicx}
\usepackage{txfonts}
\usepackage{natbib}
\begin{document}
   \title{Molecular gas in QSO host galaxies at z$>$5}


   \author{R.~Maiolino \inst{1}
          \and
          R.~Neri\inst{2}
          \and
		A.~Beelen\inst{3}
          \and
		F.~Bertoldi\inst{4}
          \and
		C.L.~Carilli\inst{5}
          \and
		P.~Caselli\inst{6,7}
          \and
		P.~Cox\inst{2}
          \and
		K.M.~Menten\inst{8}
          \and
		T.~Nagao\inst{6,9}
          \and
		A.~Omont\inst{10}
          \and
		C.M.~Walmsley\inst{6}
          \and
		F.~Walter\inst{11}
			\and
		A.~Wei\ss\inst{8}
          }

   \offprints{R. Maiolino}

   \institute{INAF - Osservatorio Astronomico di Roma, via di Frascati 33,
    00040 Monte Porzio Catone, Italy
         \and
   IRAM, 300 Rue de la Piscine, F-38406, St-Marin-d'H\`eres, France
		\and
	Institut d'Astrophysique Spatiale, Universit\'e Paris-Sud, F-91405, Orsay, France
		\and
		Argelander-Institut f\"ur Astronomie, University of Bonn,
			Auf dem Hugel 71, 53121 Bonn, Germany
		\and
	National Radio Astronomy Observatory, P.O. Box O,
	Socorro, NM 87801, USA
	 \and
	INAF - Osservatorio Astrofisico di Arcetri,
      L.go E. Fermi 5, I-50125 Firenze, Italy
      \and
		Harvard-Smithsonian Center for Astrophysics, 60 Garden Street, MS 42,
	Cambridge, MA 02138, USA
	 \and
		Max-Planck-Institut f\"ur Radioastronomie, Auf dem H\"ugel 69,
		D-53121 Bonn, Germany
         \and
	National Astronomical Observatory of Japan, 2-21-1 Osawa, 
	Mitaka, Tokyo 181-8588, Japan
         \and
	 Institut d'Astrophysique de Paris, 
	 Universit\'e Pierre \& Marie Curie, 98 bis Boulevard Arago, 
	 F-75014 Paris, France
         \and
	Max-Planck-Institut f\"ur Astronomie, K\"onigstuhl 17,
     D-69117 Heidelberg, Germany
             }

   \date{Received ; accepted }

  \abstract
   {We present observations with the IRAM Plateau de Bure Interferometer
   of three QSOs at z$>$5 aimed at detecting molecular gas in their host galaxies
   as traced by CO transitions. CO (5--4) is detected in \object{SDSS~J033829.31+002156.3} at
   z=5.0267, placing it amongst the most distant sources detected in CO.
   The CO emission is unresolved with a beam size
   of $\rm \sim 1''$, implying that the molecular gas
   is contained within a compact region, less than $\rm \sim 3~kpc$ in radius.
   We infer an upper limit on the dynamical mass of the CO emitting region
   of $\rm \sim 3\times 10^{10}M_{\sun}/\sin(i)^2$.
   The comparison with the Black Hole mass inferred from near-IR
   data suggests that the BH--to--bulge mass ratio in this galaxy is significantly higher than in
   local galaxies.
   From the CO luminosity we infer a mass reservoir of molecular gas as high as
   $\rm M(H_2)=2.4\times 10^{10}M_{\sun}$,
   implying that the molecular gas accounts for a significant fraction of the dynamical mass.
   When compared to the star formation rate derived
   from the far-IR luminosity, we infer a very short gas exhaustion timescale
   ($\sim 10^7$ years), comparable to the dynamical timescale.
   CO is not detected in the other two QSOs (\object{SDSS~J083643.85+005453.3} and 
   \object{SDSS~J163033.90+401209.6}) and upper limits are given for
   their molecular gas content. When combined with CO observations of other type 1 AGNs, spanning
   a wide redshift range (0$<$z$<$6.4), we find that the host galaxy CO luminosity (hence
   molecular gas content)
   and the AGN optical luminosity (hence BH accretion rate) are correlated,
   but the relation is not linear:
   $\rm L'_{CO}\propto [\lambda L_{\lambda}(4400\AA)]^{0.72}$. Moreover, at high redshifts (and
   especially at z$>$5) the CO luminosity appears to saturate. We discuss the implications of these
   findings in terms of black hole--galaxy co-evolution.

               }

   \maketitle
%

\section{Introduction} \label{sec_intro}

The detection of carbon monoxide (CO) emission in high redshift galaxies provides
a crucial tool for investigating the early epochs of galaxy formation
\citep[see ][for a review]{solomon05}.
Indeed, CO emission is a proxy for the molecular gas content, the reservoir
for star formation.
The CO line profile also
provides information on galaxy kinematics,
from which constraints on the dynamical mass can be inferred.

Currently only 9 galaxies 
have been detected in CO at z$>$4, only three of which are at z$>$5,
and all of them
host powerful QSOs or radio galaxies \cite{solomon05}. With the exception of the
radio-galaxy \object{TN J0924-2201} \citep{klamer05}, all of these high-z CO detections were obtained
in galaxies pre-selected amongst luminous far-IR sources (as inferred
from mm/submm continuum observations). However, such a selection criterion
may prevent us from identifying evolutionary effects in terms of molecular gas and
dust content in high-z galaxies. Indeed, in local galaxies, CO and FIR
luminosities are known to correlate \citep[e.g.][]{young91,solomon97};
hence strong far-IR emission may be
a pre-requisite for CO detection. However, at z$>$5 the ISM
is expected to undergo strong
metallicity and dust evolution, which may cause high-z galaxies to deviate from the
local CO-FIR relation. Another caveat is that present mm/submm detections are
close to the sensitivity limit of current cameras; hence even a small scatter
in the CO/FIR ratio may lead to a non-detection in $\rm L_{FIR}$.

An additional issue affecting the detection of CO in high-z QSOs is the accuracy of the redshift.
 Indeed, at z$>$4
the emission lines typically observed in the optical band 
are either strongly blueshifted with respect to the systemic velocity
of the host galaxy \citep[such as CIV at 1549\AA \ and SiIV at 1400\AA,][]{richards02}
or, in the case of
Ly$\alpha$, strongly affected by intergalactic gas absorption (Ly$\alpha$
Forest). In these cases the redshift deviations from the systemic velocity can be as large as 
several thousand $\rm km~s^{-1}$.
Until recently, millimetre receivers had bandwidths of ~0.5 GHz
(covering 1500 km~s$^{-1}$ at the best), limiting the efficiency of the search for CO in
sources with such uncertain redshift estimates. This has changed very recently
with the implementation of new receivers having a much wider bandwidth
(e.g. the new receivers at the Plateau de Bure have a bandwidth of 4 GHz).
Additionally, one can observe
lower ionization lines, such as MgII at 2798\AA \ and CIII] at 1909\AA, which
provide a better redshift estimate, since they are generally shifted by only
a few hundred $\rm km~s^{-1}$ with respect to the systemic velocity of the host galaxy.
At z$>$4 such low-ionization lines
are shifted into the near-IR, and near-IR
spectroscopic campaigns have recently provided accurate
redshifts for a number of high-z QSOs.

With the goal of increasing the number of CO detections at z$>$5, and removing any bias
towards FIR-luminous sources, we observed three high-z QSOs
with the Plateau de Bure Interferometer (PdBI).
Here we report the detection of CO emission in one source, and
upper limits for the other two.
We adopt the following cosmological parameters: $\rm
H_0=71~km~s^{-1}~Mpc^{-1}$, $\rm \Omega _{\Lambda}=0.73$ and
$\rm \Omega _m=0.27$ \citep{spergel03}.

\section{Sample selection and observations} \label{sec_obs}

The three QSOs were selected from the SDSS catalog
to be at z$>$5 and observable from the IRAM PdBI. For all, the
redshift had been re-determined with MgII or CIII] near-IR spectroscopy,
as listed in Table\ref{tab1}.
These QSOs were selected regardless of their FIR
luminosity, as inferred from previous submm/mm bolometric observations:
two of the sources have not been detected in continuum
at 850$\mu$m nor at 1.2mm (Table\ref{tab1}), while J0338+0021 has
a detection both at 850$\mu$m and at 1.2mm, from which $\rm L_{FIR}=1.5\times 10^{13}~L_{\sun}$
is inferred \citep{priddey03}. Finally, we note that J0836+0054 is 
a radio loud QSO \citep{petric03}.

Observations in configuration D
were performed between November 2005 and July 2006 with the IRAM PdBI six elements array.
The old generation 3mm receivers were tuned in single sideband to the frequency of the
redshifted CO(5--4) or CO(6--5) line, depending on the specific
redshift of each
source (Table~\ref{tab1}). The beam size in D configuration at such frequencies
is typically $\sim 5''$. The on-source integration times were 11.2 hours
for J0338$+$0021, 10.5 hours for J0836$+$0054 and 12.6 hours for J1630$+$4012.

Following the CO detection in J0338$+$0021 (see next section), this source was also observed in
configuration A (with the new PdBI receivers). Observations were obtained
in January and February 2007, for a total of 5.9~hours on source. The resulting synthesized beam
size is $\rm 1.59''\times 0.76''$ (position angle: 28$\degr$).

The resulting 1$\sigma$
sensitivities are 0.23, 0.34 and 0.26 mJy/beam in channels of 400~km~s$^{-1}$ for
J0338$+$0021, J0836$+$0054 and J1630$+$4012, respectively.

\section{Results} \label{sec_res}

\begin{figure}[!h]
   \centering
   \includegraphics[angle=-90,width=9truecm]{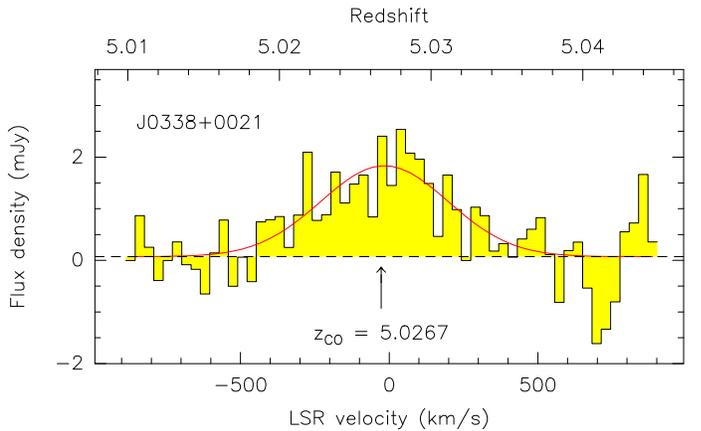}
   \caption{CO(5--4) spectrum of SDSSJ0338+0021 (sum of A and D conf.)
   rebinned to 10~MHz
   (31.4~$\rm km~s^{-1}$). The red line shows a single gaussian fit to the line.
   Velocities are relative to the MgII redshift.}
              \label{fig_spec}%
    \end{figure}

\begin{figure}[!h]
   \centering
   \includegraphics[angle=-90,width=9truecm]{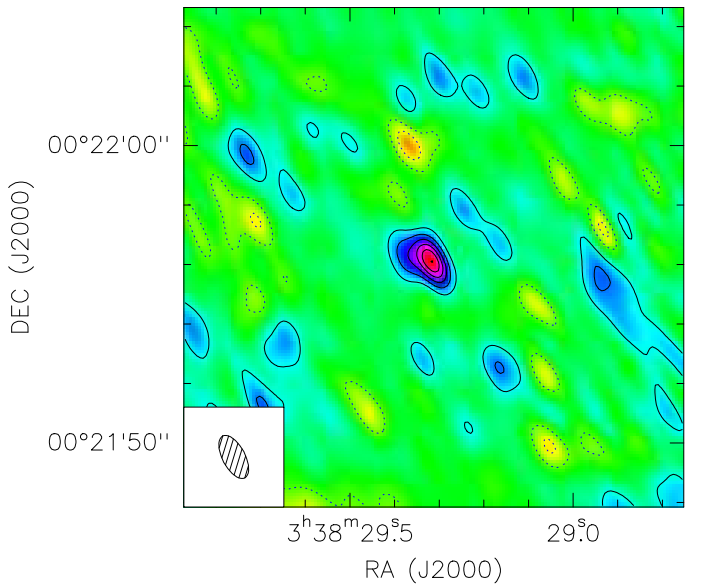}
   \caption{CO(5--4) cleaned
   map of SDSSJ0338+0021 (from both configurations A and D) obtained by integrating in
   velocity from $-$300 to $+$230~$\rm km~s^{-1}$ with contours in steps of and starting
   at $\rm 1\sigma = 0.14~Jy~beam^{-1}~km~s^{-1}$ .}
              \label{fig_map}%
    \end{figure}

CO(5--4) is detected in J0338$+$0021 with a significance
of 8$\sigma$, at a frequency of 95.6191~GHz.
This is amongst the most distant CO detections obtained so far, 
together with \object{SDSS~J114816.64+525150.3} at z=6.4, TN~J0924-2201
at z=5.2 and \object{SDSS~J092721.82+200123.7}
at z=5.77 \citep{walter03,bertoldi03a,klamer05,carilli07}.
Fig.~\ref{fig_spec} shows the spectrum and Fig.\ref{fig_map} presents the integrated
intensity map.
The rms per channel of the spectrum in Fig.~\ref{fig_spec} is 0.55~mJy/beam. We note
that the individual spectra taken in A and D configuration are consistent
with each other. The redshift inferred
from the CO line is z$=$5.0267$\pm$0.0003,
i.e. consistent with the MgII redshift (Table\ref{tab1})
within the uncertainty of $\sim$20~km~s$^{-1}$ on the line center.
The absolute position of the CO source is at RA(J200)$=$03:38:29.32 and
DEC(J2000)$=$00:21:56.1 (accuracy $\rm <0.1''$), which is consistent with the optical position.
The source is spatially unresolved, implying a radius smaller
than $\sim$2.5~kpc (along the beam minor axis).

In Table~\ref{tab1} we also report the CO luminosity $\rm L'_{CO}$
defined as in \cite{solomon97}.
The millimetric continuum at the location of J0338$+$0021 is undetected
($\rm S_{\nu}[95.6GHz]=-18\pm 85~\mu Jy$); the upper limit is
consistent with the extrapolation
of the detections at higher frequencies \citep[using the grey body fitting curves with
T$\sim$50--65K in][]{priddey03}.

Both J0836$+$0054 and J1630$+$4012 (i.e. the two FIR-faint QSOs) were undetected
in CO, and Table~\ref{tab1} gives the inferred upper limits on the CO intensity
and luminosity (assuming a line width of 400~$\rm km~s^{-1}$).

\begin{table*}[!h]
\caption{Summary of physical properties and of the CO observations for the
QSOs in our sample.}
\label{tab1}
{\centering
\begin{tabular}{lccccccccccc}
\hline\hline                 
Name & $\rm z_{MgII/CIII]}$ & $\rm S_{\nu}(850\mu m)$ & $\rm S_{\nu}(1.2mm)$ & $\rm L_{FIR}$ & CO &
   $\rm I_{CO}$ & $\rm z_{CO}$ & $\rm
FWHM$ & $\rm L'_{CO}$ & $\rm M(H_2)$ \\
     & & mJy & mJy & $\rm 10^{13} L_{\sun}$ & trans. &
	 $\rm Jy~km~s^{-1}$ & & $\rm km~s^{-1}$ &
	$\rm K~km~s^{-1}pc^2$ & $\rm 10^{10} M_{\sun}$ & \\
\hline
J033829.31+002156.3& 5.027$^a$ &11.9$\pm$2.0$^b$ &  3.7$\pm$0.3$^h$ & $\rm 1.5^b$ & (5--4) & 0.73$\pm$0.09 & 5.0267 & 500 & 
	$\rm 2.7\times 10^{10}$ & $\rm 2.2\times 10^{10}$ \\
J083643.5+005453.3 & 5.774$^c$ & 1.7$\pm$1.5$^b$ & -0.4$\pm$1.0$^g$ & $\rm <0.4^b$ & (5--4) & $<$0.41 & -- & -- &
    $\rm <1.9\times 10^{10}$ & $\rm <1.5\times 10^{10}$  \\
J163033.90+401209.6& 6.065$^d$ & 2.7$\pm$1.9$^e$ & 0.8$\pm$0.6$^f$  & $\rm <0.8^e$ & (6--5) & $<$0.30 & -- & -- &
    $\rm <1.0\times 10^{10}$ & $\rm <0.8\times 10^{10}$  \\
\hline

\hline\hline 
\end{tabular}
}
Notes: upper limits are at 3$\sigma$; references for 
the redshift inferred from the low-ionization UV lines, for the 850$\mu$m flux and
and far-IR luminosity (corrected for our adopted cosmological
constants) are: $^a$ Marinoni et al. (in prep.), $^b$ \cite{priddey03},
$^c$ \cite{stern03}, $^d$ \cite{iwamuro04}, $^e$ \cite{robson04}, $^f$ \cite{bertoldi03b},
$^g$ \cite{petric03}, $^h$ \cite{carilli01}.

\end{table*}

\section{Discussion}

\subsection{Gas and dynamical mass} \label{sec_galaxy_mass}

The CO luminosity can be used to infer
the molecular gas content. Most authors adopt the conversion factor
$\rm \alpha = 0.8~M_{\sun}~(K~km~s^{-1}~pc^2)^{-1}$
between CO(1--0) line luminosity and $\rm M(H_2)$, as inferred for nearby
starbursts with moderate CO excitation and in virial equilibrium \citep{downes98,
solomon05}.
Such a conversion
factor may not be appropriate for high-z QSOs, which often show indications of
high gas excitation \citep{bertoldi03a,weiss07}. We do not have information
on the CO transitions lower than (5--4) in J0338$+$0021, and hence we
cannot constrain the gas excitation. Given these uncertainties, we assume
the same conversion factor as for local starbursts, and we also assume constant
brightness temperature (the optically thick case) from J=1 to J=5, i.e.
$\rm L'_{CO}(1-0)=L'_{CO}(5-4)$. Under these assumptions we infer
a molecular gas mass of $\rm \sim 2.2\times 10^{10}~M_{\sun}$. Note that
this is probably a lower limit on the H$_2$ mass, since the
transition (5--4) is likely subthermal and the conversion factor probably higher,
as inferred in other starburst galaxies and powerful sources
\citep[e.g.][]{bayet06,bradford03,weiss07}.

The line width and the upper limit on the CO size of J0338$+$0021 allow us to infer
an upper limit on the dynamical mass. Following \cite{tacconi06} we derive
$\rm v_c\sin (i)$ (where $\rm v_c$ is the circular velocity at the outer CO
radius, and ``i'' the inclination angle of the gaseous disk) by
dividing the CO line FWHM by 2.4. We obtain
$\rm M_{dyn} = R v_c^2/G < 3.2\times 10^{10}M_{\sun}/\sin^2(i)$,
where we have assumed an upper limit for the size of the CO source of 1$''$
($\sim$ average of our beam sizes), hence R$<$3.2~kpc.
The main uncertainty of the dynamical mass upper limit is due to the unknown
inclination angle ``i''. As discussed in \cite{carilli06}, type 1 AGNs may be
biased against edge-on host disks, because in such cases the nucleus should be obscured.
Such a bias is inferred from the finding that the CO emission in (type 1) QSOs is systematically
narrower than in SMGs at similar redshifts.
However, we note that in the specific case of J0338$+$0021 the CO FWHM is amongst the
largest ever observed in QSOs \citep[whose median FWHM is
300~km~s$^{-1}$,][]{carilli06}, and similar to the median FWHM observed
in SMGs (500~km~s$^{-1}$), suggesting that J0338$+$0021 is
likely observed at high inclination.

Based on these results we conclude that the molecular gas mass accounts for a substantial
fraction of the dynamical mass. More specifically, if the system
is nearly edge-on ($\rm i\sim 90\degr$) then the molecular gas mass accounts for more than
70\% of the dynamical mass. Even if the system has an inclination of
$\rm i = 30\degr$ the molecular gas mass still accounts for more than 20\% of the
dynamical mass. Such large fractions of molecular gas mass are also observed in local
ULIRGs \citep{sanders96} as well as in distant SMGs \citep{tacconi06}, and indicate that
the host galaxy of J0338$+$0021 is in an early evolutionary stage.

\subsection{The black hole--bulge mass ratio at z$\sim$5} \label{sec_bhbulge}

Based on the width of the MgII~2798\AA \ line
and the continuum intensity at $\rm \lambda _{rest}=3000\AA$ (Marinoni et al. in prep.),
and by following the prescription in \cite{mclure02},
we estimate a black hole mass in this QSO of $\rm M_{BH}\approx 2.5\times 10^8 M_{\sun}$.
We can infer an upper limit on the mass of a putative stellar bulge, by using the upper limit
on the dynamical mass obtained above and subtracting the molecular gas mass
(and assuming that the bulge is smaller than 3.5~kpc in radius).
If the molecular gas disk is nearly edge-on, we derive
$\rm M_{BH}/M_{bulge}>2.5\times 10^{-2}$, which is
substantially larger than the ratio observed locally, i.e. $\sim 10^{-3}$
\citep{marconi03}\footnote{The value of $\rm M_{BH}/M_{bulge}\sim 0.002$ given in
\cite{marconi03} has to be lowered by a factor of 5/3 to account for more recent
estimates of the bulge virial masses (Marconi priv. comm.).}.
In order to have the lower limit
on $\rm M_{BH}/M_{bulge}$ marginally
consistent with the local value, the inclination of the gas disk in J0338$+$0021 must be about
20$\degr$, i.e. close to the average value found by \cite{carilli06} for other QSOs with CO
detection. However, as discussed above, the very broad CO emission of J0338$+$0021 relative
to other QSOs suggests that the gaseous disk in the former is much more inclined. Moreover,
one should keep in mind that the inferred $\rm M_{BH}/M_{bulge}$ is a {\it  lower limit}.
It is difficult to obtain more quantitative constraints without higher resolution and higher
sensitivity data. However, the current observations suggest that the $\rm M_{BH}/M_{bulge}$
at high-z is higher than observed locally. This result is in agreement with 
the $\rm M_{BH}/M_{bulge}$ ratio inferred for the most distant QSO J1148+5251 at z$=$6.4
by \cite{walter04}. Other indications of a higher, with respect to local,
$\rm M_{BH}/M_{bulge}$ mass ratio
were found by \cite{peng06} and \cite{mclure06} in z$>$1 AGNs. All of these results suggest
that BH growth occurred on timescales shorter
than bulge formation, and that the locally observed
BH-bulge relation was achieved only at z$<$1.

\subsection{Star formation efficiency} \label{sec_sfe}

The far-IR emission is regarded as a tracer of the star formation rate
\citep{kennicutt98}. In QSOs the possible contamination by dust heated by the
AGN has been a hotly debated issue; however
recent observations have shown that at least in the {\it far-}IR,
the emission is generally due to star formation even
in the case of powerful QSOs \citep{schweitzer06,lutz07,wang07}.
In J0338$+$0021 the observed $\rm L_{FIR}=1.5\times 10^{13}~L_{\sun}$ implies
a star formation rate of $\rm \sim 2500~M_{\sun}~yr^{-1}$, if using
the $\rm L_{FIR}$--to--SFR conversion factor derived by \cite{kennicutt98}.
The ratio
$\rm L_{FIR}/L'_{CO}$ is considered a measure of the star formation
efficiency, since it is related to the star formation rate per unit of molecular
gas mass \citep{solomon05}.
$\rm L_{FIR}/L'_{CO}$ is found to steadily increase with
luminosity \citep[e.g.][]{solomon97}, which is interpreted as an
increasing star formation efficiency in the most powerful starburst systems.
J0338$+$0021
has a very high $\rm L_{FIR}/L'_{CO}$ ratio, implying
very high star formation efficiency. More specifically, in J0338$+$0021,
the whole molecular gas content is expected to be converted into stars on a
time scale of only $\rm \sim 10^7~yrs$, i.e. a few times the dynamical timescale within
the CO radius ($\rm t_{dyn}=R/v_c<1.5\times10^7~yr$). Cases like J0338$+$0021 are rare but not
unique; indeed similar ``maximal starburst'' systems are found among other
hyper-luminous infrared galaxies at lower redshift, both QSOs and
starbursts \citep[e.g.][]{tacconi06}.

   \begin{figure}[!h]
   \centering
   \includegraphics[width=9.3truecm]{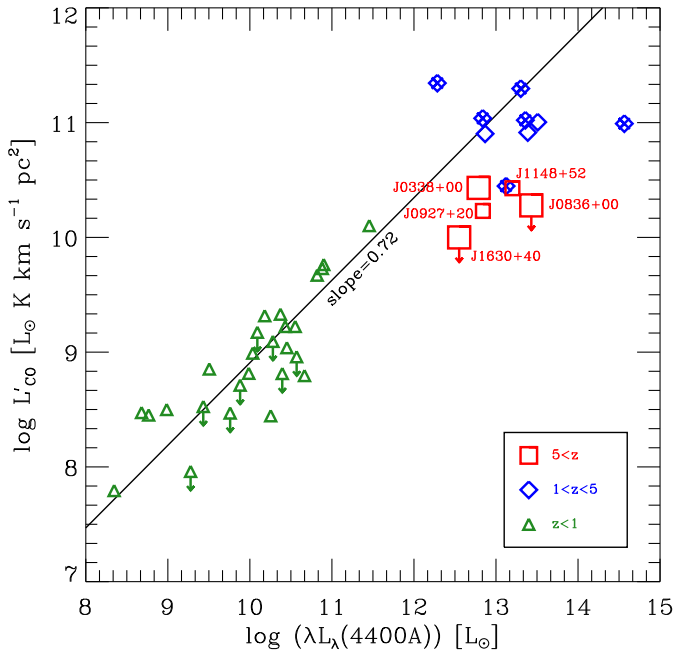}
   \caption{$\rm L'_{CO}$ versus (nuclear) $\lambda L_{\lambda}(4400\AA )$
    for type 1 AGNs with CO measurements. Green triangles are sources at z$<$1,
	blue diamonds are sources at 1$<$z$<$5, red squares are sources at z$>$5.
	Large squares indicate the new measurements at z$>$5 reported in this paper.
   Crosses mark strongly lensed QSOs, for which differential magnification
   of the nuclear optical emission and CO emission in the host galaxy may occur.
   The black solid line indicates a linear fit to the data at z$<$1. Note the
   non-linearity of the relation and the likely flattening of the relationship at
   high redshift and high optical luminosities.}
              \label{fig_coopt}%
    \end{figure}

\subsection{Black hole accretion and molecular gas reservoir} \label{sec_coopt}

If we focus on type 1 (unobscured) AGNs, it is interesting to compare the
molecular gas content, as traced by $\rm L'_{CO}$, with the optical luminosity,
the latter being proportional to the Black Hole accretion rate.
For type 1 AGNs with CO measurements \citep[from][]{sanders91,evans06,solomon97,solomon05,
maiolino97},
we have derived the rest-frame optical luminosity
($\rm \lambda L_{\lambda}(4400\AA )$)
by using spectroscopic or photometric data available from the literature
at the observed wavelength closest to (1+z)4400\AA \ (generally near-IR
data for QSOs at high redshift). For nearby Sy1 we considered only objects
with nuclear measurements of the optical flux, to minimize the stellar light
contamination.
Fig.~\ref{fig_coopt} shows $\rm L'_{CO}$ as a function of
$\rm \lambda L_{\lambda}(4400\AA )$, where sources in different redshift
ranges are identified by different symbols and colors.
Objects marked by a cross are strongly lensed QSOs: in these cases differential magnification
may occur between nuclear optical emission and CO emission in the host
galaxy\footnote{Note that high angular resolution observations
rule out significant lensing for all of the QSOs at z$>$5 considered here
\citep{richards06,richards04,frey05,white05}, with exception of J1630$+$4012
for which no high angular resolution data are available.}.
CO and optical luminosity clearly correlate, but the relation is non-linear.
More specifically, by fitting the low-z data (z$<$1) alone, we obtain
\begin{equation}
\rm \frac{L'_{CO}}{L_{\odot}~K~km~s^{-1}~pc^2}=
51.8\times \left [ \frac{\lambda L_{\lambda}(4400\AA)}{L_{\odot}} \right ] ^{0.72\pm
0.08}.
\end{equation}
High-z
QSOs show indications for further flattening of the
relation at high luminosities. In particular, QSOs at z$>$5 hint at a possible
saturation of the CO luminosity at a few times $\rm 10^{10}~L_{\odot}~km~s^{-1}~pc^{2}$.

The correlation between BH accretion rate
($\rm \lambda L_{\lambda}(4400\AA)$) and molecular gas content in the host galaxy
($\rm L'_{CO}$) may be at the origin of the correlation between
BH mass and stellar bulge mass observed in local galaxies \citep[e.g. ][]{ferrarese00,marconi03}.
The non--linearity of the relation is probably a consequence of the fact that, while BH accretion
can be arbitrarily high (limited only by the Eddington luminosity), the molecular gas content is
limited by the galaxy evolutionary processes. In particular, the saturation of the CO luminosity
in sources at z$>$5 may indicate that galaxy evolutionary mechanisms cannot assemble more
than a few times $\rm 10^{10}~M_{\odot}$ of gas in such early evolutionary phases of
galaxy formation. Such results are consistent with the finding
of a saturation also in terms of star formation rate in high-z, luminous QSOs
\citep{maiolino07}.

Regardless of their interpretation, these results may explain the low galaxy--to--BH mass
ratio observed in high-z QSOs (especially at z=6.4),
relative to local galaxies (Sect.~\ref{sec_bhbulge}).

\subsection{Low FIR/CO ratios are not common amongst high-z, luminous sources} \label{sec_firco}

The detection of CO (5--4) and (1--0) in the far-IR weak (submm undetected) radio galaxy
TNJ0924-2201 at z=5.2 by \cite{klamer05}, suggested the possible existence
of a significant population of high-z sources with large reservoirs of molecular
gas but with little dust emission. These could be cases where dust had little
time to form, or whose average dust temperature is extremely cold. However, the
non-detection of CO in the two far-IR faint QSOs in our sample does not provide
additional support for the existence of a large population of such objects, and
TNJ0924-2201 remains the only case of exceptionally low
$\rm L_{FIR}/L_{CO}$ (about a factor of 5 lower than sources with similar
CO luminosity). As a consequence, strong far-IR emission seems to generally be a
prerequisite for CO detection at high redshift.
Of course the statistics are still extremely poor, and more
observations are required to investigate this. Moreover, we 
cannot rule out the possibility
that in the two QSOs without CO detection the low-ionization
UV lines provide
a redshift which is offset by more than 1000~$\rm km~s^{-1}$ (which would move
the CO line out of the old 3mm receiver band),
although MgII has a velocity generally consistent with the systemic velocity
within at most a few hundred $\rm km~s^{-1}$ \citep{vandenberk01,richards02},
and our result on J0338$+$0021 supports this scenario.

\begin{acknowledgements}
      RM, PC, and MW acknowledge support from INAF.
	  We thank the IRAM staff for their support during the observations.
    IRAM is supported by INSU/CNRS (France), MPG (Germany) and IGN (Spain).
\end{acknowledgements}


\begin{thebibliography}{}

\bibitem[Bayet et al.(2006)]{bayet06} Bayet, E., Gerin, M., 
Phillips, T.~G., \& Contursi, A.\ 2006, \aap, 460, 467

\bibitem[Bertoldi et al.(2003)]{bertoldi03a} Bertoldi, F., et al.\ 
2003, \aap, 409, L47

\bibitem[Bertoldi et al.(2003)]{bertoldi03b} Bertoldi, F.,
et al.\ 2003, \aap, 406, L55

\bibitem[Bradford et al.(2003)]{bradford03} Bradford, C.~M., 
Nikola, T., Stacey, G.~J., et al.\ 2003, \apj, 586, 891

\bibitem[Carilli et al.(2001)]{carilli01} Carilli, C.~L., et al.\ 
2001, \apj, 555, 625 

\bibitem[Carilli \& Wang(2006)]{carilli06} Carilli, C.~L., \& 
Wang, R.\ 2006, \apj, 131, 2763

\bibitem[Carilli et al.(2007)]{carilli07} Carilli, C.~L.,
et al.\ 2007, \apj, in press

\bibitem[Downes \& Solomon(1998)]{downes98} Downes, D., \& 
Solomon, P.~M.\ 1998, \apj, 507, 615

\bibitem[Evans et al.(2006)]{evans06} Evans, A.~S., et al.\ 2006, \aj, 132, 2398

\bibitem[Ferrarese \& Merritt(2000)]{ferrarese00} Ferrarese, L., \& 
Merritt, D.\ 2000, \apjl, 539, L9

\bibitem[Frey et al.(2005)]{frey05} Frey, S., Paragi, Z., 
Mosoni, L., \& Gurvits, L.~I.\ 2005, \aap, 436, L13

\bibitem[Granato et al.(2004)]{granato04} Granato, G.~L., et
al.\ 2004, \apj, 600, 580


\bibitem[Iwamuro et al.(2004)]{iwamuro04} Iwamuro, F., Kimura, 
M., et al.\ 2004, \apj, 
614, 69

\bibitem[Kennicutt(1998)]{kennicutt98} Kennicutt, R.~C., Jr.\ 1998,
  \araa, 36, 189

\bibitem[Klamer et al.(2005)]{klamer05} Klamer, I.~J., Ekers, 
R.~D., et al.\ 2005, 
\apjl, 621, L1

\bibitem[Lutz et al.(2007)]{lutz07} Lutz, D., et al.\ 
2007, \apj, in press, arXiv:0704.0133v1 [astro-ph]


\bibitem[Maiolino et al.(1997)]{maiolino97} Maiolino, R., Ruiz, 
M., Rieke, G.~H., \& Papadopoulos, P.\ 1997, \apj, 485, 552

\bibitem[Maiolino et al.(2007)]{maiolino07} Maiolino, R., Shemmer, 
O., Imanishi, M., Netzer, H., Oliva, E., Lutz, D., \& Sturm, E.\ 2007,
\aap, 468, 979

\bibitem[Marconi \& Hunt(2003)]{marconi03} Marconi, A., \& Hunt, 
L.~K.\ 2003, \apjl, 589, L21

\bibitem[McLure \& Jarvis(2002)]{mclure02} McLure, R.~J., \& 
Jarvis, M.~J.\ 2002, \mnras, 337, 109

\bibitem[McLure et al.(2006)]{mclure06} McLure, R.~J., Jarvis, M.~J.,
  Targett, T.~A., Dunlop, J.~S., \& Best, P.~N.\ 2006, \mnras, 368,
  1395




\bibitem[Peng et al.(2006)]{peng06} Peng, C.~Y., et al.\ 2006, \apj, 640, 114

\bibitem[Petric et al.(2003)]{petric03} Petric, A.~O., Carilli, 
C.~L., Bertoldi, F., et al.\ 2003, \aj, 126, 15

\bibitem[Priddey et al.(2003)]{priddey03} Priddey, R.~S., Isaak, 
K.~G., McMahon, R.~G., Robson, E.~I., \& Pearson, C.~P.\ 2003, \mnras, 344, 
L74

\bibitem[Richards et al.(2002)]{richards02} Richards, G.~T., 
Vanden Berk, D.~E., Reichard, T.~A., et al.\ 2002, \aj, 124, 1

\bibitem[Richards et al.(2004)]{richards04} Richards, G.~T., et 
al.\ 2004, \aj, 127, 1305

\bibitem[Richards et al.(2006)]{richards06} Richards, G.~T., et 
al.\ 2006, \aj, 131, 49


\bibitem[Robson et al.(2004)]{robson04} Robson, I.,
et al.\ 2004, \mnras, 351, L29

\bibitem[Sanders et al.(1991)]{sanders91} Sanders, D.~B., 
Scoville, N.~Z., \& Soifer, B.~T.\ 1991, \apj, 370, 158

\bibitem[Sanders \& Mirabel(1996)]{sanders96} Sanders, D.~B., \& 
Mirabel, I.~F.\ 1996, \araa, 34, 749

\bibitem[Schweitzer et al.(2006)]{schweitzer06} Schweitzer, M., et 
al.\ 2006, \apj, 649, 79

\bibitem[Solomon et al.(1997)]{solomon97} Solomon, P.~M.,
et al.\ 1997, \apj, 478, 144

\bibitem[Solomon \& Vanden Bout(2005)]{solomon05} Solomon, P.~M., 
\& Vanden Bout, P.~A.\ 2005, \araa, 43, 677

\bibitem[Spergel et al.(2003)]{spergel03} Spergel, D.~N., et al.\
  2003, \apjs, 148, 175

\bibitem[Stern et al.(2003)]{stern03} Stern, D., Hall, P.~B., 
Barrientos, L.~F., et al.\ 2003, \apjl, 596, L39

\bibitem[Tacconi et al.(2006)]{tacconi06} Tacconi, L.~J., et al.\ 
2006, \apj, 640, 228

\bibitem[Vanden Berk et al.(2001)]{vandenberk01} Vanden Berk, D.~E., 
et al.\ 2001, \aj, 122, 549

\bibitem[Young \& Scoville(1991)]{young91} Young, J.~S., \& 
Scoville, N.~Z.\ 1991, \araa, 29, 581

\bibitem[Walter et al.(2003)]{walter03} Walter, F., et al.\ 
2003, \nat, 424, 406

\bibitem[Walter et al.(2004)]{walter04} Walter, F., Carilli, C., 
Bertoldi, et al.\ 
2004, \apjl, 615, L17

\bibitem[Wang et al.(2007)]{wang07} Wang, R., et al.\ 
2007, \apj, in press, arXiv:0704.2053v1 [astro-ph]

\bibitem[Wei{\ss} et al.(2007)]{weiss07} Wei\ss, A., et al., \aap, in
press (astro-ph/0702669)


\bibitem[White et al.(2005)]{white05} White, R.~L., Becker, 
R.~H., Fan, X., \& Strauss, M.~A.\ 2005, \aj, 129, 2102

\end{thebibliography}
\end{document}